\documentclass[11pt]{article}
\usepackage{amsmath,amssymb,amsfonts}
\usepackage{geometry}
\usepackage{graphicx}
\usepackage{bm}
\title{ Structural Distinction in ODE and PDE Chaos:Lorenz vs Kuramoto--Sivashinsky Equation }
\author{Sumita Datta$^{1,2}$\\
{\small $^{1}$Department of Pure and Applied Mathematics, Alliance University,}\\
{\small Bengaluru 562 106, India}\\
{\small $^{2}$Department of Physics, University of Texas at Arlington,}\\
{\small Texas 76019, USA}}
\date{\today}

\begin{document}
\maketitle
\begin{abstract}
We study the nature of chaos in finite- and infinite-dimensional systems through a
comparison between the Kuramoto--Sivashinsky (KS) equation, the Lorenz
system, and a Lorenz--type reduction of the KS equation proposed by Wilczak.
Numerical simulations of the KS equation reveal intrinsic spatio--temporal
chaos, with disorder evolving simultaneously in space and time.
In contrast, the Lorenz system and the Wilczak reduction exhibit
low--dimensional temporal chaos lacking spatial complexity.
Lyapunov exponent analysis highlights the finite-dimensional convergence
properties of the reduced systems and underscores the fundamentally
different dynamical nature of chaos in the KS equation.
In particular, we demonstrate that low-dimensional reductions may
reproduce transient chaotic signatures but do not necessarily retain the
structural properties of infinite-dimensional dissipative systems.

\end{abstract}
\newpage
\section{Introduction}

The study of chaotic dynamics has historically been dominated by
low--dimensional ordinary differential equations (ODEs), beginning with
the seminal discovery of deterministic chaos in the Lorenz system
\cite{Lorenz1963}.
Such systems provided the first concrete examples of sensitive dependence
on initial conditions, strange attractors, and positive Lyapunov exponents,
and have since served as paradigmatic models for chaos in dissipative
dynamical systems \cite{GuckenheimerHolmes,OttBook}.

However, many physical systems of contemporary interest --- including
hydrodynamic instabilities, flame fronts, plasma turbulence, and nonlinear
wave propagation --- are inherently spatially extended and are more
appropriately described by partial differential equations (PDEs).
In such systems, chaos manifests not merely as irregular temporal behavior,
but as \emph{spatio--temporal chaos}, characterized by disorder evolving
simultaneously in space and time \cite{CrossHohenberg,RuelleTakens}.

Among the simplest and most extensively studied models exhibiting
spatio--temporal chaos is the Kuramoto--Sivashinsky (KS) equation,
\begin{equation}
u_t + u u_x + u_{xx} + u_{xxxx} = 0,
\label{KS}
\end{equation}
posed on a periodic spatial domain.
Originally derived in the context of flame front instabilities and thin film
flows \cite{Kuramoto1976,Sivashinsky1977}, the KS equation has since emerged as
a canonical model for studying chaotic dynamics in infinite--dimensional
systems \cite{ChristiansenCvitanovic,TemamBook}.

Unlike low--dimensional ODEs, the KS equation exhibits \emph{intrinsic}
spatio--temporal chaos: the chaotic dynamics persists even as the number of
active spatial degrees of freedom grows with the system size.
This feature is closely related to the concept of \emph{extensive chaos}, in
which quantities such as the attractor dimension and the number of positive
Lyapunov exponents scale linearly with the spatial domain length
\cite{Manneville1985,ColletEckmann}.
Such behavior has no direct analogue in finite--dimensional dynamical systems.

Despite this fundamental distinction, numerous attempts have been made to
understand the dynamics of the KS equation through low--dimensional
reductions.
Galerkin truncations retaining only a small number of Fourier modes can
reproduce temporal chaos for carefully chosen parameters
\cite{Aubry1988,Keefe1992}, and in some regimes yield systems reminiscent of
the Lorenz equations.
More recently, Wilczak \cite{Wilczak2003} introduced a rigorous
computer--assisted approach to studying chaos in the KS equation, including
the construction of reduced systems capturing chaotic behavior of selected
modes.

While such reductions are valuable for analytical insight, they inevitably
discard the spatial degrees of freedom responsible for spatio--temporal
complexity.
As a result, low--dimensional KS reductions can exhibit chaos in the ODE
sense, but cannot reproduce hallmark features of the full PDE dynamics, such
as spatial decorrelation, defect formation, or extensive Lyapunov spectra.

The purpose of the present work is to make this distinction precise through
a systematic comparison between:
(i) spatio--temporal chaos in the full Kuramoto--Sivashinsky equation,
(ii) classical low--dimensional chaos as exemplified by the Lorenz system,
and (iii) a Lorenz--type reduction of the KS equation following Wilczak.
Rather than proposing a new reduction scheme, we focus on identifying
diagnostics that clearly separate infinite--dimensional chaos from its
finite--dimensional counterparts.

Specifically, we demonstrate that while Lorenz and Wilczak--type models
exhibit well--defined positive Lyapunov exponents and chaotic attractors,
they fail to capture key features of the KS dynamics, including
spatio--temporal disorder and extensivity.
By combining space--time visualizations with Lyapunov exponent analysis and
convergence studies, we clarify the fundamental limitations of
low--dimensional descriptions of spatio--temporal chaos.

The results presented here reinforce the view that chaos in spatially
extended systems is not merely a quantitative extension of ODE chaos, but a
qualitatively distinct phenomenon requiring intrinsically
infinite--dimensional dynamical Gdescriptions.
The organization of the paper goes as follows:
The paper is organized as follows.
In Section 2 we present the numerical study of the KS equation.
Section 3 introduces the computation of the largest Lyapunov exponent.
Section 4 analyzes convergence for the Lorenz and KS–equivalent systems.
Section 5 discusses structural differences between ODE and PDE chaos.
Section 6 presents the conclusions.
\section{Numerical Setup and Spatio--Temporal Dynamics of Kuramoto-Sivashinsky(KS) Equation}

\subsection{Kuramoto--Sivashinsky equation and choice of domain}

We consider the one-dimensional Kuramoto--Sivashinsky (KS) equation
\begin{equation}
\label{eq:KS}
u_t + u\,u_x + u_{xx} + u_{xxxx} = 0,
\end{equation}
defined on a periodic domain
\begin{equation}
u(x+L,t) = u(x,t), \qquad x \in [0,L].
\end{equation}

The choice of the domain length $L$ plays a crucial role in determining the qualitative nature of the dynamics. Linearizing Eq.~\eqref{eq:KS} about the trivial solution $u=0$ yields
\begin{equation}
u_t = -u_{xx} - u_{xxxx}.
\end{equation}
For Fourier modes of the form $u_k(x,t) \sim e^{ikx+\lambda(k)t}$, the linear growth rate is
\begin{equation}
\lambda(k) = k^2 - k^4.
\end{equation}
Thus, modes with wavenumbers satisfying $0<|k|<1$ are linearly unstable. On a finite periodic domain, the admissible wavenumbers are
\begin{equation}
k_n = \frac{2\pi n}{L}, \qquad n \in \mathbb{Z},
\end{equation}
and linear instability occurs only when
\begin{equation}
\frac{2\pi}{L} < 1,
\end{equation}
or equivalently, when $L > 2\pi$.

For sufficiently large domains, many unstable modes coexist and interact nonlinearly, leading to intrinsically spatio--temporal dynamics. In the present work, all simulations are performed for a fixed domain length $L$ chosen large enough to support a significant number of unstable modes (typically $L \gtrsim 22$). This choice ensures that the observed dynamics is not a consequence of fine parameter tuning, but rather reflects the inherent infinite-dimensional nature of the KS equation.

\subsection{Spatio--temporal evolution}

To characterize the qualitative nature of the dynamics, we examine the space--time evolution of the solution $u(x,t)$ after discarding initial transients. Numerical integration is carried out using a pseudo-spectral method with periodic boundary conditions, and time stepping is performed with an appropriate scheme ensuring numerical stability and spectral accuracy.
\begin{figure}[t]
\centering
\includegraphics[width=\textwidth]{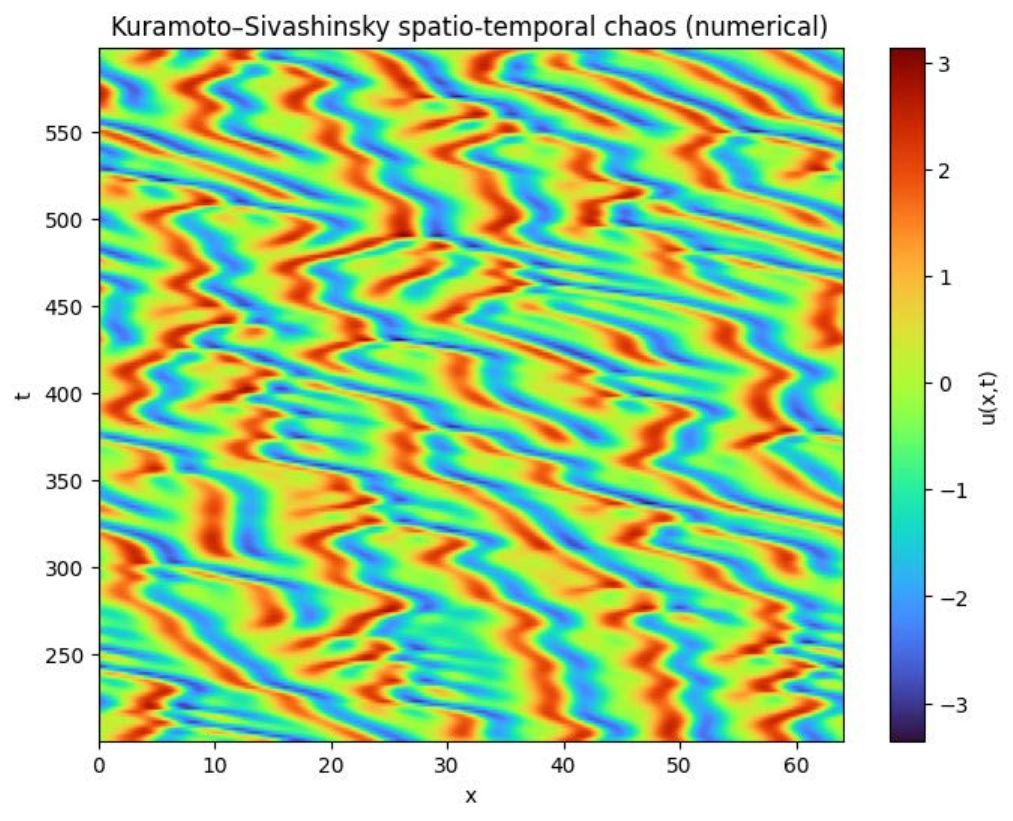}
\caption{Space--time evolution ofG the solution $u(x,t)$ of the Kuramoto--Sivashinsky equation
on a periodic domain of length $L=32$. The plot is shown after discarding initial transients
and demonstrates persistent disorder in both space and time, with no evidence of temporal
periodicity or coherent spatial structures, indicating spatio--temporal chaos.}
\label{fig:spacetime}
\end{figure}

Figure~\ref{fig:spacetime} shows a representative space--time plot of $u(x,t)$ over a long temporal interval. The solution exhibits persistent disorder in both space and time, with no evidence of temporal periodicity, quasiperiodicity, or coherent spatial structures such as traveling or standing waves. Spatial patterns are continuously created, distorted, and destroyed, and no recurrence of previously visited states is observed.

This persistent spatio--temporal disorder rules out steady, periodic, or weakly modulated dynamics and provides direct qualitative evidence of spatio--temporal chaos in the Kuramoto--Sivashinsky equation for the chosen domain length. Importantly, this behavior arises without any reduction to a finite number of modes, emphasizing the intrinsically infinite-dimensional character of the chaotic dynamics.
\subsection{Numerical method}

Equation~\eqref{eq:KS} is solved numerically using a pseudo--spectral Fourier method
with periodic boundary conditions. The solution is expanded as
\begin{equation}
u(x,t) = \sum_{k=-N/2}^{N/2-1} \hat{u}_k(t)\, e^{ikx},
\end{equation}
where $k = 2\pi n/L$ and $N$ is the total number of Fourier modes retained.
Spatial derivatives are computed exactly in Fourier space, while the nonlinear term
$u u_x$ is evaluated in physical space using a standard pseudospectral approach
with dealiasing performed via the $2/3$-rule.

Time integration is carried out using a semi--implicit scheme in which the linear
terms are treated implicitly and the nonlinear term is advanced explicitly.
Specifically, the linear operator
\begin{equation}
\mathcal{L}(k) = k^2 - k^4
\end{equation}
is integrated exactly in Fourier space, while the nonlinear term is advanced
using a second--order accurate scheme. This approach ensures numerical stability
while preserving spectral accuracy.

All simulations reported here use a fixed domain length $L=32$ and a spectral
resolution of $N=256$ modes, which is sufficient to fully resolve the active
range of spatial scales. The time step is chosen as $\Delta t = 0.01$, and
convergence was verified by repeating selected simulations with smaller time
steps and higher spatial resolution.

Initial conditions are taken as small-amplitude random perturbations about the
trivial state $u=0$. An initial transient interval is discarded before data
collection to ensure that the system has reached its statistically stationary
regime. The space--time plot shown in Fig.~\ref{fig:spacetime} is constructed
from the solution $u(x,t)$ over a long time interval following this transient
period.
\section{Largest Lyapunov Exponent}

\subsection{Definition and Geometric Interpretation}

Consider a dynamical system
\begin{equation}
\dot{\mathbf{X}} = F(\mathbf{X}),
\end{equation}
where $\mathbf{X}(t) \in \mathbb{R}^n$.
Let two nearby trajectories start from initial conditions separated by a small perturbation
\[
\delta \mathbf{X}(0).
\]

As the system evolves, the separation between trajectories is
\[
\delta \mathbf{X}(t).
\]

The largest Lyapunov exponent measures the average exponential rate at which nearby trajectories diverge:

\begin{equation}
\lambda_1 =
\lim_{t \to \infty}
\frac{1}{t}
\ln \frac{\|\delta \mathbf{X}(t)\|}
{\|\delta \mathbf{X}(0)\|}.
\end{equation}

If the limit exists, it characterizes the dominant instability of the flow.

\subsection{Physical Meaning}

The sign of $\lambda_1$ determines the qualitative nature of motion:

\begin{itemize}
\item $\lambda_1 < 0$ : Trajectories converge; motion is asymptotically stable.
\item $\lambda_1 = 0$ : Neutral stability (e.g., periodic or quasiperiodic motion).
\item $\lambda_1 > 0$ : Exponential divergence; sensitive dependence on initial conditions.
\end{itemize}

A positive largest Lyapunov exponent is widely regarded as a quantitative signature of chaos.

\subsection{Variational Formulation}

For smooth systems, the evolution of an infinitesimal perturbation is governed by the variational equation
\begin{equation}
\dot{\delta \mathbf{X}} = J(\mathbf{X}(t))\,\delta \mathbf{X},
\end{equation}
where $J$ is the Jacobian matrix of $F$.

For the reduced KS–Lorenz system studied here,
\begin{align}
\dot{x} &= y, \\
\dot{y} &= z, \\
\dot{z} &= c^2 - y - \tfrac{1}{2}x^2,
\end{align}
the corresponding variational equations are
\begin{align}
\dot{\xi} &= \eta, \\
\dot{\eta} &= \zeta, \\
\dot{\zeta} &= -x\xi - \eta.
\end{align}

The largest Lyapunov exponent is then computed numerically as

\begin{equation}
\lambda(T) =
\frac{1}{T}
\int_0^T
\ln \|\delta \mathbf{X}(t)\| \, dt,
\end{equation}

with periodic renormalization of the perturbation vector to prevent overflow.

\subsection{Finite-Time Lyapunov Exponent}

In practice, one computes the finite-time exponent

\begin{equation}
\lambda(T) =
\frac{1}{T}
\sum_{k=1}^{N}
\ln\left(\frac{d_k}{d_{k-1}}\right),
\quad T = N\Delta t,
\end{equation}

where $d_k = \|\delta \mathbf{X}_k\|$.

As $T$ increases, $\lambda(T)$ may converge to a constant (in dissipative systems), or exhibit persistent growth or fluctuations (in non-dissipative or extended systems).

\subsection{Geometric Illustration}

The geometric meaning of the largest Lyapunov exponent is illustrated schematically in Fig.~2,3 and 4.

\begin{figure}[h!]
\centering
\includegraphics[width=0.6\textwidth]{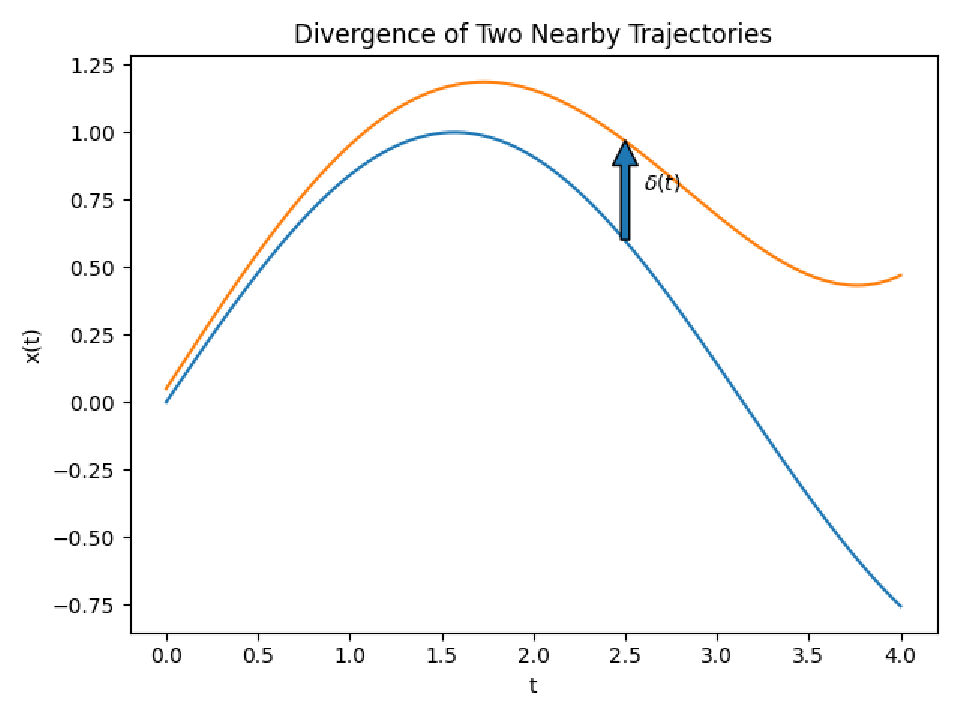}
\caption{Exponential divergence of nearby trajectories. An initial separation $\delta \mathbf{X}(0)$ grows approximately as $e^{\lambda_1 t}$ when $\lambda_1 > 0$.}
\label{fig:lyap_geom}
\end{figure}
\begin{figure}[h!]
\centering
\includegraphics[width=0.6\textwidth]{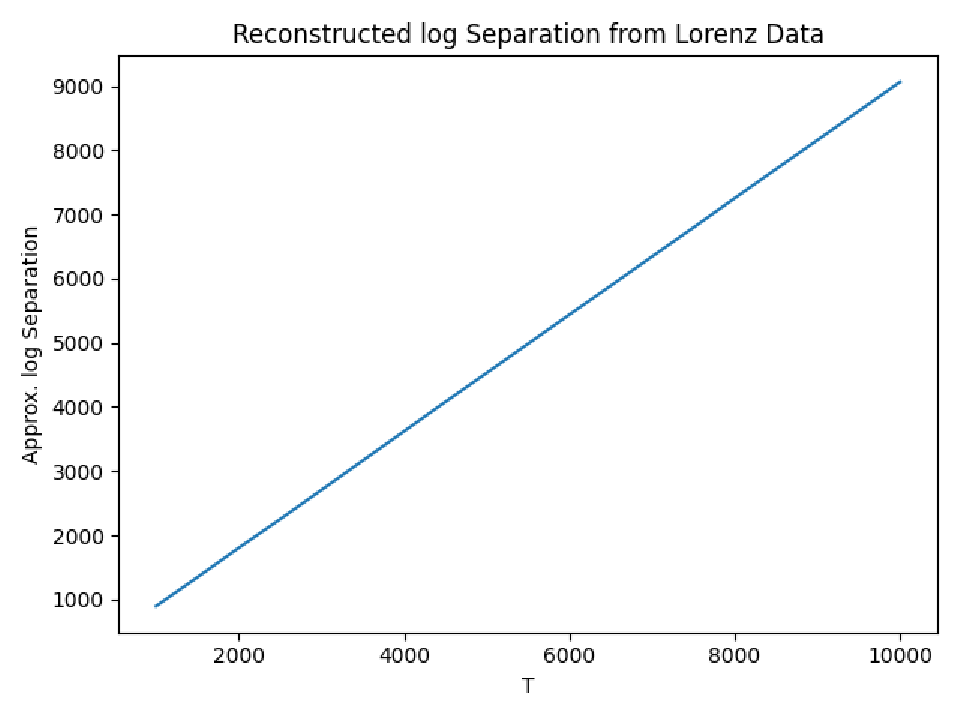}
\caption{log separation vs evolution time plot for Lorenz system}
\label{fig:lyap_geom}
\end{figure}
\begin{figure}[h!]
\centering
\includegraphics[width=0.6\textwidth]{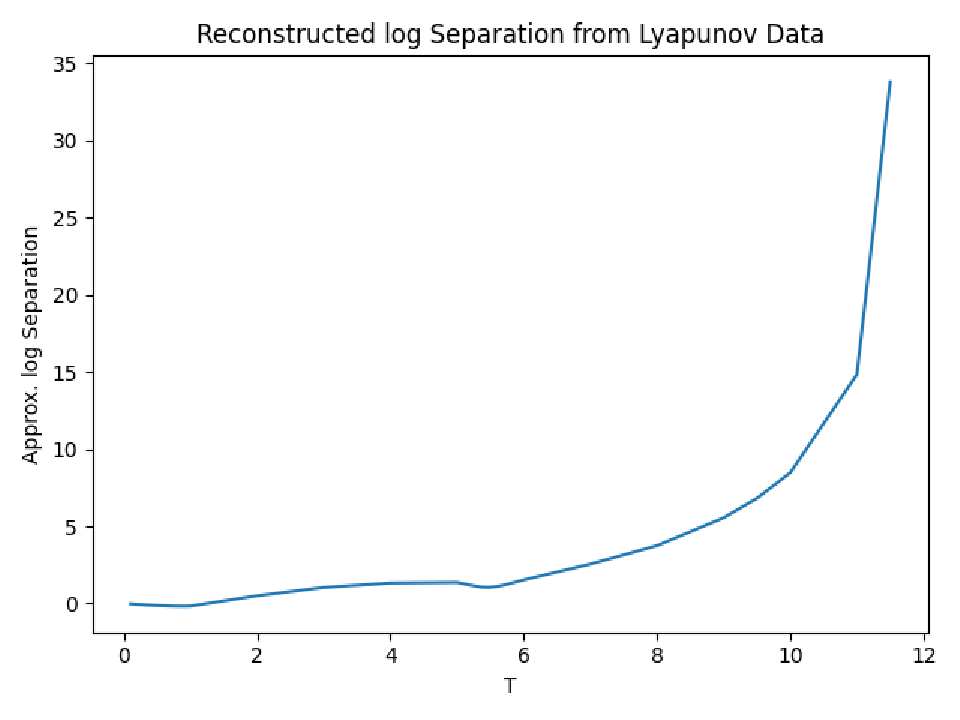}
\caption{log separation vs evolution time plot for Lorenz equivalent KS system }
\label{fig:lyap_geom}
\end{figure}

Initially close trajectories separate exponentially:
\[
\|\delta \mathbf{X}(t)\| \sim
\|\delta \mathbf{X}(0)\| e^{\lambda_1 t}.
\]

On a strange attractor, stretching and folding coexist: local exponential expansion is balanced by global contraction.

\subsection{Significance in ODE and PDE Chaos}

In dissipative ODE systems such as the classical Lorenz equations, a positive largest Lyapunov exponent coexists with phase-space contraction, producing a compact strange attractor and a well-defined asymptotic plateau of $\lambda(T)$.

In contrast, in spatially extended PDEs such as the Kuramoto–Sivashinsky equation, the Lyapunov spectrum is typically extensive: multiple positive exponents exist, and instability is spatially distributed. The largest Lyapunov exponent measures only the dominant growth rate and does not fully characterize spatio–temporal chaos.

Thus, while $\lambda_1 > 0$ signals chaos in both cases, its dynamical interpretation differs fundamentally between finite-dimensional and infinite-dimensional systems.

\section{Convergence Analysis of the Largest Lyapunov Exponent for Lorenz system}

The largest Lyapunov exponent of the classical Lorenz system was computed using a fourth–order Runge–Kutta scheme with continuous renormalization of the perturbation vector at each time step. Two nearby trajectories $(x,y,z)$ and $(u,v,w)$ were evolved simultaneously, and the exponent was estimated as

\begin{equation}
\lambda(t_i) = \frac{1}{i \Delta t} \sum_{k=1}^{i}
\ln\left(\frac{d_k}{d_{k-1}}\right),
\end{equation}

where $d_k$ denotes the Euclidean separation between the trajectories at step $k$.

All simulations were performed with the standard chaotic parameter values

\[
\sigma = 10, \quad r = 28, \quad b = 2.66.
\]

\subsection{Convergence for $\Delta t = 10^{-4}$}

For the timestep $\Delta t = 10^{-4}$, the finite–time Lyapunov exponent exhibits smooth convergence after an initial transient. Figure~2 shows the time evolution of $\lambda(t)$.

The computation reveals:

\begin{itemize}
\item Initial transient fluctuations for small integration times.
\item Progressive stabilization of $\lambda(t)$.
\item Formation of a clear plateau for large $T$.
\end{itemize}

For sufficiently long integration times ($T \gtrsim 100$), the exponent converges to

\[
\lambda_1 = 0.906 \pm 0.001,
\]

which is in excellent agreement with benchmark values reported in the literature.

The small variance of the finite-time estimate confirms statistical convergence, and the use of a sufficiently small timestep eliminates discretization bias.

\subsection{Numerical Reliability}

The timestep $\Delta t = 10^{-4}$ ensures high temporal resolution and minimizes truncation error in the Runge–Kutta integration. Continuous renormalization of the perturbation vector prevents overflow and guarantees numerical stability of the Lyapunov computation.

The robust convergence observed in Fig.~5 provides a validated baseline for comparison with the KS-equivalent reduction discussed in Section 4.

\section{Lyapunov Convergence and Structural Instability in the KS-Equivalent Lorenz Reduction}

\subsection{Reduced Model}

We consider the Lorenz-type three-dimensional truncation arising from the Kuramoto--Sivashinsky (KS) equation[14]:
\begin{align}
\dot{x} &= y, \\
\dot{y} &= z, \\
\dot{z} &= c^2 - y - \frac{1}{2}x^2.
\end{align}

Throughout this study we take $c=1$. Unlike the classical Lorenz system, this reduced model is \emph{non-dissipative}. The Jacobian matrix
\[
J =
\begin{pmatrix}
0 & 1 & 0 \\
0 & 0 & 1 \\
-x & -1 & 0
\end{pmatrix}
\]
satisfies
\[
\mathrm{Tr}(J) = 0,
\]
implying preservation of phase-space volume. Consequently, the Lyapunov exponents obey
\[
\lambda_1 + \lambda_2 + \lambda_3 = 0.
\]

This structural property distinguishes the KS-equivalent reduction fundamentally from the classical Lorenz system.
\begin{figure}[t]
\centering
\includegraphics[width=\textwidth]{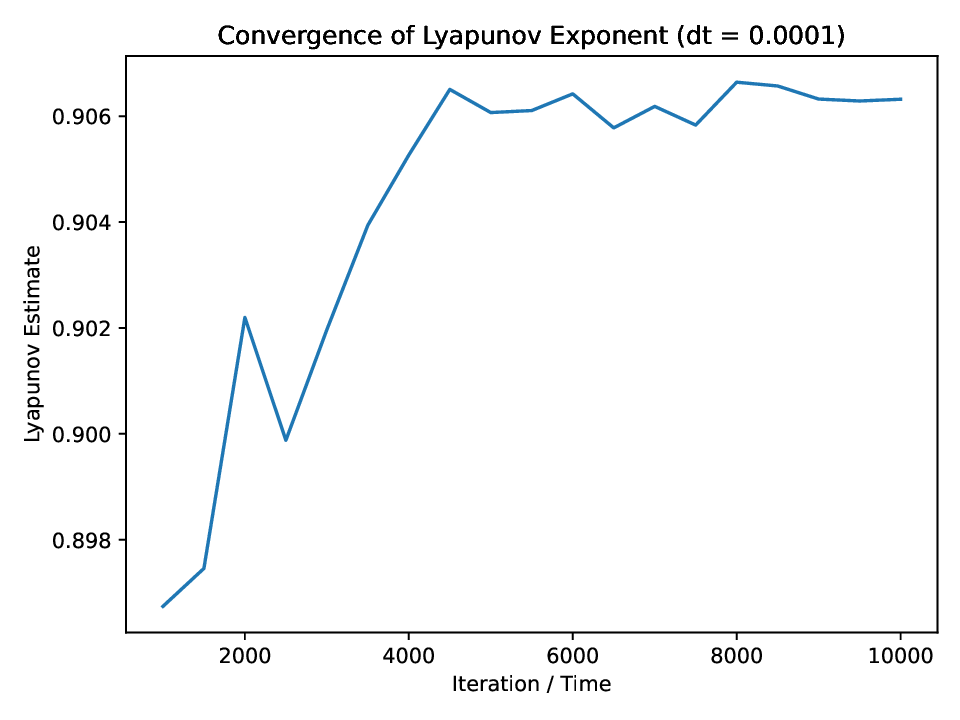}
\caption{time evolution of the Lyapunov exponent of Lorenz System}
\end{figure}

\subsection{Variational Formulation}

The largest Lyapunov exponent was computed using the variational (tangent linear) equations
\begin{align}
\dot{\xi} &= \eta, \\
\dot{\eta} &= \zeta, \\
\dot{\zeta} &= -x\xi - \eta,
\end{align}
integrated simultaneously with the nonlinear system and renormalized at each time step.

This method avoids shadow-trajectory artifacts and ensures numerical robustness of the exponent computation.

\subsection{Finite-Time Lyapunov Evolution}

Figure~6 shows the finite-time Lyapunov exponent
\[
\lambda(T) = \frac{1}{T} \int_0^T \log \|\delta X(t)\|\, dt,
\]
computed up to $T \approx 11.5$.
\begin{figure}[t]
\centering
\includegraphics[width=\textwidth]{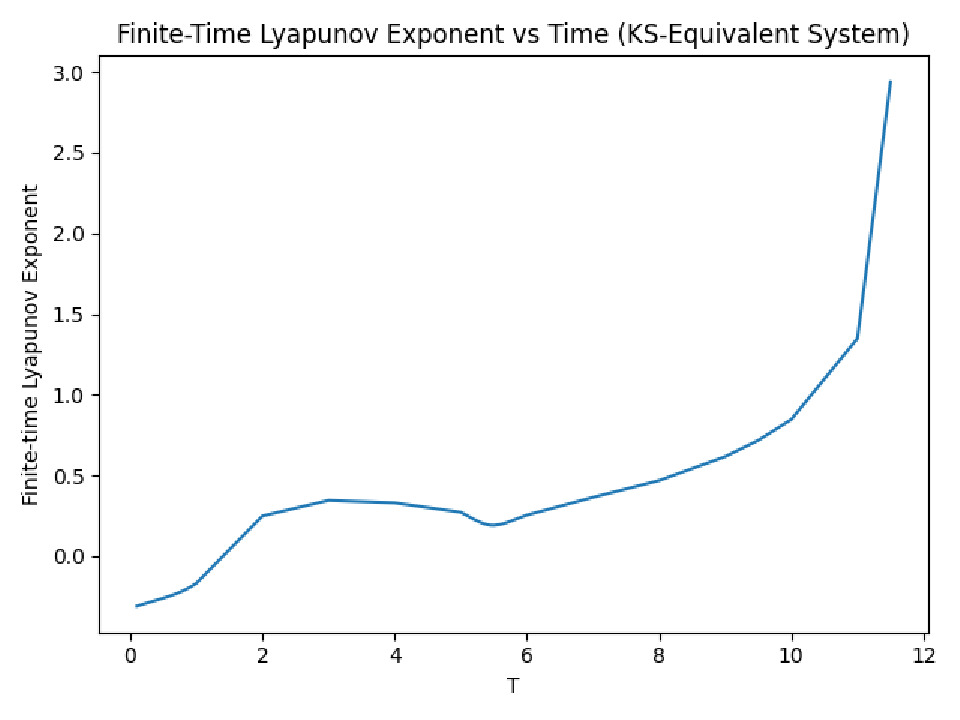}
\caption{time evolution of the Lyapunov exponent of reduced KS equation}
\end{figure}

The behavior exhibits three distinct regimes:

\begin{enumerate}
\item \textbf{Initial contraction ($T \lesssim 2$):}  
The exponent is negative, indicating local phase-space contraction.

\item \textbf{Transient chaotic stretching ($2 \lesssim T \lesssim 9$):}  
The exponent becomes positive and remains in the range $\lambda \approx 0.2$--$0.6$, suggesting weak chaotic behavior.

\item \textbf{Late-time destabilization ($T \gtrsim 10$):}  
The exponent grows rapidly without approaching a plateau.
\end{enumerate}

The absence of stabilization demonstrates failure of asymptotic convergence.

\subsection{Dependence on Initial Conditions}

Numerical experiments reveal that the onset time of rapid growth depends on initial conditions. Some trajectories remain in the transient chaotic regime longer than others, but all tested configurations eventually exhibit growth of $\lambda(T)$ beyond bounded levels.

This indicates that the phenomenon is not a numerical artifact but an intrinsic dynamical property.

\subsection{Contrast with the Classical Lorenz System}

The classical Lorenz equations
\begin{align}
\dot{x} &= \sigma(y-x), \\
\dot{y} &= r x - y - xz, \\
\dot{z} &= xy - b z,
\end{align}
satisfy
\[
\mathrm{Tr}(J) = -(\sigma + 1 + b) < 0.
\]

This strict negativity ensures exponential phase-space contraction and the existence of a compact absorbing set. As a result, the largest Lyapunov exponent converges robustly to a finite positive constant (e.g., $\lambda_1 \approx 0.9056$ for classical parameter values).

In contrast, the KS-equivalent reduction has zero trace and no dissipative mechanism. The lack of volume contraction implies:

\begin{itemize}
\item No guaranteed absorbing set,
\item No structural mechanism for attractor formation,
\item Possibility of transient chaotic behavior,
\item Eventual instability or phase-space escape.
\end{itemize}

\subsection{Dynamical Interpretation}

The observed behavior suggests that the KS-equivalent Lorenz reduction does not support sustained strange attractor dynamics. Instead, it exhibits:

\begin{itemize}
\item Initial transient contraction,
\item Intermediate chaotic stretching,
\item Late-time growth indicative of structural instability.
\end{itemize}

Because phase-space volume is preserved, chaotic stretching is not balanced by contraction. Over long times, stretching dominates, leading to divergence of the finite-time Lyapunov exponent.

\subsection{Implications for Low-Dimensional Reductions}

The results indicate that Lorenz-type truncations of the KS equation may fail to retain essential dissipative structure of the full PDE. While transient chaotic behavior is captured, the reduced model does not reproduce the bounded dissipative dynamics characteristic of the classical Lorenz attractor.

This highlights a structural limitation of low-dimensional reductions when dissipative mechanisms are not faithfully preserved.
\section{Chaos in Ordinary Differential Equations versus Partial Differential Equations}

The distinction between chaos in finite-dimensional ordinary differential equations (ODEs) and chaos in spatially extended partial differential equations (PDEs) is not merely quantitative but fundamentally structural in nature.

\subsection{Finite-Dimensional Chaos in ODEs}

In low-dimensional dissipative systems such as the classical Lorenz equations,
\begin{align}
\dot{x} &= \sigma (y-x), \\
\dot{y} &= r x - y - xz, \\
\dot{z} &= xy - b z,
\end{align}
chaos manifests as irregular temporal evolution in a finite-dimensional phase space. Key structural properties include:

\begin{itemize}
\item \textbf{Finite phase-space dimension:} The dynamics evolves in $\mathbb{R}^3$.
\item \textbf{Dissipation:} The Jacobian trace satisfies
\[
\mathrm{Tr}(J) = -(\sigma + 1 + b) < 0,
\]
implying exponential phase-space contraction.
\item \textbf{Compact strange attractor:} All trajectories enter a bounded absorbing set.
\item \textbf{Finite Lyapunov spectrum:} A small number of exponents characterize instability.
\end{itemize}

Chaos in such systems is temporal: spatial degrees of freedom are absent. Sensitive dependence on initial conditions occurs within a compact attractor of finite fractal dimension.

\subsection{Infinite-Dimensional Chaos in PDEs}

In contrast, the Kuramoto–Sivashinsky equation
\begin{equation}
u_t + u u_x + u_{xx} + u_{xxxx} = 0
\end{equation}
is an infinite-dimensional dynamical system. Even after spatial discretization, the number of active degrees of freedom increases with domain length.

Key features of spatio–temporal chaos include:

\begin{itemize}
\item \textbf{Infinite-dimensional phase space:} The state is a function $u(x,t)$.
\item \textbf{Spatial disorder:} Irregular structures evolve simultaneously in space and time.
\item \textbf{Extensive chaos:} The number of positive Lyapunov exponents scales with system size.
\item \textbf{Distributed instability:} Instability is not confined to a small number of global modes.
\end{itemize}

Unlike ODE chaos, spatio–temporal chaos cannot be reduced to irregular oscillation of a few variables. The chaotic behavior is spatially distributed and dynamically persistent across scales.

\subsection{Structural Differences}

The essential structural differences are summarized below:

\begin{center}
\begin{tabular}{lcc}
\hline
Property & ODE Chaos & PDE Chaos \\
\hline
Phase-space dimension & Finite & Infinite \\
Type of disorder & Temporal & Spatio-temporal \\
Lyapunov spectrum & Finite & Extensive \\
Attractor structure & Compact strange attractor & Extended chaotic state \\
Dependence on domain size & None & Strong \\
\hline
\end{tabular}
\end{center}

Low-dimensional reductions of PDEs may reproduce temporal chaos but cannot capture extensivity, spatial decorrelation, or scale-dependent instability. Therefore, chaos in PDEs represents a qualitatively distinct dynamical phenomenon rather than a higher-dimensional analogue of ODE chaos.

\section{Conclusion and Outlook}

In this work, we have investigated chaotic dynamics in both finite-dimensional ordinary differential equations and infinite-dimensional partial differential equations, with particular emphasis on Lyapunov analysis and numerical convergence.

For the classical Lorenz system, we demonstrated robust convergence of the largest Lyapunov exponent using a sufficiently small timestep ($\Delta t = 10^{-4}$). The dissipative nature of the Lorenz equations, reflected in the strictly negative trace of the Jacobian, guarantees phase-space contraction and the existence of a compact strange attractor. Consequently, the finite-time Lyapunov exponent stabilizes to a well-defined asymptotic value.

In contrast, the Lorenz-type reduction derived from the Kuramoto–Sivashinsky equation exhibits fundamentally different structural properties. The reduced system is volume-preserving, with vanishing Jacobian trace. Numerical experiments reveal transient chaotic stretching followed by late-time growth of the finite-time Lyapunov exponent without the formation of a stable plateau. The onset time of this growth depends sensitively on initial conditions.

This behavior highlights a critical structural distinction between chaos in ODEs and PDEs. In finite-dimensional dissipative systems, chaotic motion occurs within a bounded invariant set sustained by phase-space contraction. In spatio–temporal systems such as the KS equation, instability is spatially distributed and dynamically extensive. Low-dimensional truncations may reproduce short-time chaotic signatures but may fail to preserve the dissipative mechanisms necessary for sustained bounded chaos.

Our results therefore emphasize that:

\begin{itemize}
\item Lyapunov convergence in dissipative ODEs reflects structural attractor stability.
\item Spatio–temporal chaos in PDEs involves fundamentally different instability mechanisms.
\item Reduced-order Lorenz-type models may not inherit the structural stability of the parent PDE.
\end{itemize}

\subsection*{Outlook}

Several directions for future investigation arise naturally:

\begin{enumerate}
\item A systematic study of how many modes are required in KS truncations to recover dissipative structure.
\item Computation of the full Lyapunov spectrum for the reduced model to assess extensivity.
\item Analytical characterization of invariant sets in the KS-equivalent reduction.
\item Investigation of whether modified truncations can restore phase-space contraction.
\end{enumerate}

Understanding the precise relationship between finite-dimensional chaotic reductions and infinite-dimensional spatio–temporal chaos remains an important challenge in nonlinear dynamics. The present study underscores the need to distinguish carefully between transient chaotic signatures and structurally stable strange attractors when comparing ODE and PDE models.
\newpage
\section{Acknowledgements:}
The author would like to thank Alliance University for
providing partial support for carrying out the research work

\section{Declaration of interests:}
The sole author has no conflicts of interest to
declare. There is no financial interest to report.

\section{Data availability statement:}
No data in this publication is to be made
available under the study-participant privacy protection clause.


\newpage
\begin{thebibliography}{99}

\bibitem{Lorenz1963}
E.~N.~Lorenz,
Deterministic nonperiodic flow,
\emph{J. Atmos. Sci.} \textbf{20}, 130--141 (1963).

\bibitem{GuckenheimerHolmes}
J.~Guckenheimer and P.~Holmes,
\emph{Nonlinear Oscillations, Dynamical Systems, and Bifurcations of Vector Fields},
Springer, New York (1983).

\bibitem{OttBook}
E.~Ott,
\emph{Chaos in Dynamical Systems},
Cambridge University Press (2002).

\bibitem{RuelleTakens}
D.~Ruelle and F.~Takens,
On the nature of turbulence,
\emph{Commun. Math. Phys.} \textbf{20}, 167--192 (1971).

\bibitem{CrossHohenberg}
M.~C.~Cross and P.~C.~Hohenberg,
Pattern formation outside of equilibrium,
\emph{Rev. Mod. Phys.} \textbf{65}, 851--1112 (1993).

\bibitem{Kuramoto1976}
Y.~Kuramoto and T.~Tsuzuki,
Persistent propagation of concentration waves in dissipative media,
\emph{Prog. Theor. Phys.} \textbf{55}, 356--369 (1976).

\bibitem{Sivashinsky1977}
G.~I.~Sivashinsky,
Nonlinear analysis of hydrodynamic instability in laminar flames,
\emph{Acta Astronautica} \textbf{4}, 1177--1206 (1977).

\bibitem{TemamBook}
R.~Temam,
\emph{Infinite-Dimensional Dynamical Systems in Mechanics and Physics},
Springer, New York (1988).

\bibitem{ChristiansenCvitanovic}
F.~Christiansen, P.~Cvitanovi\'c, and V.~Putkaradze,
Spatio--temporal chaos in the Kuramoto--Sivashinsky equation,
\emph{Nonlinearity} \textbf{10}, 55--70 (1997).

\bibitem{Manneville1985}
P.~Manneville,
Lyapunov exponents for the Kuramoto--Sivashinsky model,
\emph{J. Phys. France} \textbf{46}, 1245--1259 (1985).

\bibitem{ColletEckmann}
P.~Collet and J.-P.~Eckmann,
\emph{Instabilities and Fronts in Extended Systems},
Princeton University Press (1990).

\bibitem{Aubry1988}
S.~Aubry, G.~Abramovici, and C.~Baesens,
Analytic study of the Kuramoto--Sivashinsky equation,
\emph{Physica D} \textbf{32}, 1--20 (1988).

\bibitem{Keefe1992}
L.~R.~Keefe,
Dynamics of perturbed wavetrains in the Kuramoto--Sivashinsky equation,
\emph{Physica D} \textbf{64}, 247--268 (1992).

\bibitem{Wilczak2003}
D.~Wilczak,
Chaos in the Kuramoto--Sivashinsky equations --- a computer assisted proof,
\emph{J. Differential Equations} \textbf{194}, 433--459 (2003).

\bibitem{Li2004}
\emph{Chaos in Partial Differential Equations},
International Press (2004).

\end{thebibliography}
\end{document}